\documentclass[aps,prl,twocolumn,groupedaddress,showpacs]{revtex4}

\usepackage{bm}
\usepackage{graphicx}

\begin{document}

\title{Manipulation of Topological States and Bulk Band Gap Using Natural Heterostructures of a Topological Insulator}

\author{
	K. Nakayama,$^{1,\ast}$
	K. Eto,$^2$
	Y. Tanaka,$^1$
	T. Sato,$^{1,\S}$
	S. Souma,$^3$
	T. Takahashi,$^{1,3}$
	Kouji Segawa,$^2$
	and Yoichi Ando$^{2,\dagger}$}

\affiliation{$^1$Department of Physics, Tohoku University, Sendai 980-8578, Japan}
\affiliation{$^2$Institute of Scientific and Industrial Research, Osaka University, Ibaraki, Osaka 567-0047, Japan}
\affiliation{$^3$WPI Research Center, Advanced Institute for Materials Research, Tohoku University, Sendai 980-8577, Japan}

\date{\today}

\begin{abstract}
We have performed angle-resolved photoemission spectroscopy on (PbSe)$_5$(Bi$_2$Se$_3$)$_{3m}$, which forms a natural multilayer heterostructure consisting of a topological insulator (TI) and an ordinary insulator. For $m$ = 2, we observed a gapped Dirac-cone state within the bulk-band gap, suggesting that the topological interface states are effectively encapsulated by block layers; furthermore, it was found that the quantum confinement effect of the band dispersions of Bi$_2$Se$_3$ layers enhances the effective bulk-band gap to 0.5 eV, the largest ever observed in TIs. In addition, we found that the system is no longer in the topological phase at $m$ = 1, pointing to a topological phase transition between $m$ = 1 and 2. These results demonstrate that utilization of naturally-occurring heterostructures is a new promising strategy for realizing exotic quantum phenomena and device applications of TIs.
\end{abstract}


\maketitle

Three-dimensional (3D) topological insulators (TIs) realize a topological quantum state associated with unusual metallic surface states which appear within the bulk-band gap \cite{HasanReview, ZhangReview}. The topological surface states are characterized by a Dirac-cone energy dispersion with a helical spin texture. Owing to the peculiar spin texture, the Dirac fermions in the TIs are immune to backward scattering by nonmagnetic impurities or disorder \cite{Yazdani1, Yazdani2} and carry dissipationless spin current \cite{KaneMele}, holding promise for exploring fundamental physics, spintronics, and quantum computing \cite{HasanReview, ZhangReview}. However, there are a number of challenges that need to be overcome before TIs meet those promises. For example, while experimental realizations of novel topological phenomena depend crucially on the inherent robustness of the topological surface states against perturbations, it turned out to be difficult to maintain stable surface properties under ambient atmosphere \cite{TaskinPRL, KongNChem}. Also, potential applications of TIs for a wide range of devices working at room temperature require a large bulk-band gap, but the gap value reported to date is $\sim$0.35 eV at most \cite{HasanReview, SatoTBS}. Such a situation has been a hindrance for realizing novel topological phenomena and device applications of TIs, calling for a conceptually new approach to the manipulation of materials properties of TIs.

A commonly used strategy for such a manipulation is the chemical substitution of constituent elements, as has been widely tried in systems based on Bi$_2$Se$_3$ and Bi$_2$Te$_3$ \cite{TaskinPRL, ChenBi2Te3, HsiehTunable, CavaQI, AnalytisQO, ButchPRB, RenQO}. Another, potentially more effective approach is the heterostructure engineering where one can alter the stacking sequence of layers or insert different building blocks into the crystal, which may trigger gigantic quantum effects and/or new physical phenomena. However, this method has not been seriously explored in 3D TIs owing to a limited number of TI materials discovered to date.

In this Letter, we demonstrate that utilization of naturally-occurring heterostructures in bulk crystals containing TI units is a promising pathway to overcome aforementioned problems. Specifically, we show high-resolution angle-resolved photoemission spectroscopy (ARPES) data for lead-based (Pb-based) homologous series, (PbSe)$_5$(Bi$_2$Se$_3$)$_{3m}$, which forms a natural multilayer heterostructure consisting of a TI and an ordinary insulator.  Our results suggest that quantum confinement effects due to the characteristic layer stacking of this homologous series result in the protection of the topological states and the largest bulk-band gap among known TIs. In addition, this system appears to present an intriguing topological phase transition between $m$ = 1 and 2.

In tetradymite Bi$_2$Se$_3$, the ordered Se-Bi-Se-Bi-Se quintuple layer (QL) forms the basic unit, which is stacked along the (111) direction. In the case of our Pb-based compound whose chemical composition can be expressed as [(PbSe)$_5$]$_n$[(Bi$_2$Se$_3$)$_3$]$_m$, the crystal consists of $m$ QLs of Bi$_2$Se$_3$ sandwiched by adjacent $n$ bilayers of PbSe \cite{PbCrystal, PbXray} [Fig. 1(a)], thus offering an excellent example to systematically alter the building blocks of the crystal, and is suitable for investigating new functional properties of TIs.  We performed ARPES measurements for a fixed $n$ value ($n$ = 1) with $m$ = 1 (Pb$_5$Bi$_6$Se$_{14}$) and $m$ = 2 (Pb$_5$Bi$_{12}$Se$_{23}$). The data are compared with that for Bi$_2$Se$_3$, which can be viewed as a member of this homologous series with ($n$,$m$) = (1,$\infty$).

High-quality single crystals of (PbSe)$_5$[(Bi$_2$Se$_3$)$_3$]$_m$ were grown by a modified Bridgeman method using high purity elements (Pb 99.998$\%$, Bi and Se 99.9999$\%$) in a sealed evacuated quartz tube. The phase diagram of Pb-Bi-Se ternary system is very complicated \cite{Shelimova}, which causes various crystal phases to compete during the crystal growth and one typically finds multiple phases in a boule grown by the Bridgeman method. For this experiment, the starting composition was chosen to be Pb:Bi:Se = 2:2:5, and after the growth, the dominant phase in the boule was found to be different for the top and bottom parts of the boule. We chose the targeted phase ($i.e.$ $m$ = 1 or 2) based on the x-ray diffraction analysis of the crystals cut out from the boule. Nevertheless, it turned out from the ARPES measurements that a finite amount of mixtures of different phases in the bulk crystal are unavoidable ($e.g.$, a few percent of $m$ = 2 domains are mixed in the $m$ = 1 sample as confirmed by the x-ray diffraction measurement). Hence, we focused the light beam to a small spot ($\sim$0.1 mm) and scanned the beam-spot position on the cleaved surface to make sure that the ARPES data are taken on a single domain of the desired phase. We confirmed that the ARPES data from the same domain give essentially the same result even when the nominal composition of bulk crystal is different. ARPES measurements were performed with a VG-Scienta SES2002 electron analyzer with a tunable synchrotron light at the beamline BL28A at Photon Factory (KEK).  We used circularly polarized lights of 36-60 eV. The energy and angular resolutions were set at 15-30 meV and 0.2$^{\circ}$, respectively. Samples were cleaved $in$-$situ$ along the (111) crystal plane in an ultrahigh vacuum of 1$\times$10$^{-10}$ Torr. A shiny mirror-like surface was obtained after cleaving the samples, confirming its high quality. The Fermi level ($E_{\rm F}$) of the samples was referenced to that of a gold film evaporated onto the sample holder.

\begin{figure}[!t]
\begin{center}
\includegraphics[width=3.4in]{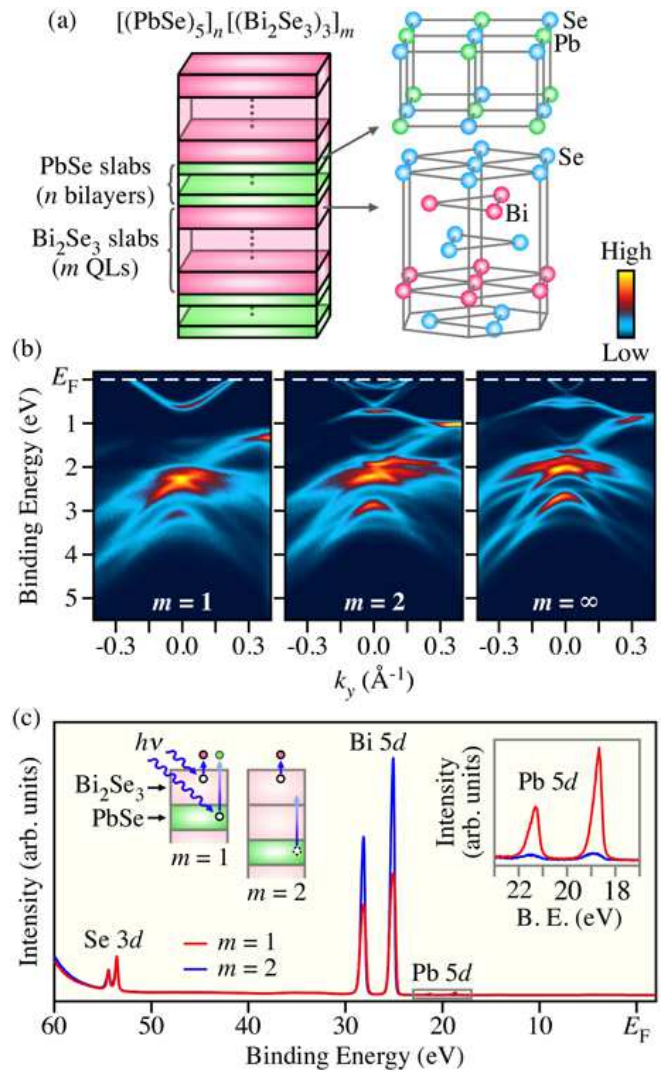}
\end{center}
\caption{
(color online). (a) Schematic illustration of the building blocks of [(PbSe)$_5$]$_n$[(Bi$_2$Se$_3$)$_3$]$_m$ homologous series. (b) Comparison of the valence-band ARPES intensities of (PbSe)$_5$(Bi$_2$Se$_3$)$_{3m}$ for $m$ = 1, 2, and $\infty$, plotted as a function of $E_{\rm B}$ and wave vector measured along the $k_y$ axis (the $\bar{\Gamma}$$\bar{\rm K}$ cut) at $T$ = 30 K with $h$$\nu$ = 60 eV. (c) Normal-emission photoemission spectra in a wide energy region for $m$ = 1 and 2 measured with $h$$\nu$ = 80 eV at $T$ = 30 K. Right inset shows a magnified view around the Pb-5$d$ core levels. Schematic illustration of the cleaved surface and the photoemission process for $m$ = 1 and 2 is also displayed in left inset.
}
\end{figure}

We first show in Fig. 1(b) the valence-band ARPES intensity maps in the binding energy ($E_{\rm B}$) $vs$ momentum ($\bm{{\rm k}}$) plane measured along the $\bar{\Gamma}\bar{\rm K}$ cut of the Brillouin zone, compared for $m$ = 1, 2, and $\infty$. They exhibit common characteristics below $E_{\rm B}$ of $\sim$1 eV, suggesting that the spectral features for $m$ = 1 and 2 are essentially dominated by the contribution from the Bi$_2$Se$_3$ layer. A closer look reveals that the overall band dispersions for $m$ = 1 and 2 are shifted downward with respect to that for $m$ = $\infty$, likely due to an increase in Se vacancies caused by Pb substitution as also reported for other Pb-based TIs \cite{SoumaPb124, PbNCommun}. The dominant contribution from the Bi$_2$Se$_3$ layer is also confirmed by the measurements of core levels. As shown in Fig. 1(c), the intensity of Pb-5$d$ core levels is much weaker than that of Bi 5$d$ in both $m$ = 1 and 2, while their photo-ionization cross-sections are similar \cite{YehLindau}. This is likely to indicate that the topmost layer of the cleaved surface is the Bi$_2$Se$_3$ QL in the present experimental condition. Moreover, the Pb-5$d$ intensity observed for $m$ = 2 is much weaker than that for $m$ = 1 (right inset), and this can be naturally understood if the top PbSe layer is located deeper beneath the surface in the $m$ = 2 case (see left inset). In fact, a rough estimate that takes into account the finite photoelectron escape length (0.5 nm) and the depth of the topmost PbSe layer (1 and 2 nm for $m$ = 1 and 2, respectively) suggests an order of magnitude weaker Pb-5$d$ intensity for $m$ = 2, which is in line with the present experimental result.

\begin{figure}[!t]
\begin{center}
\includegraphics[width=3.4in]{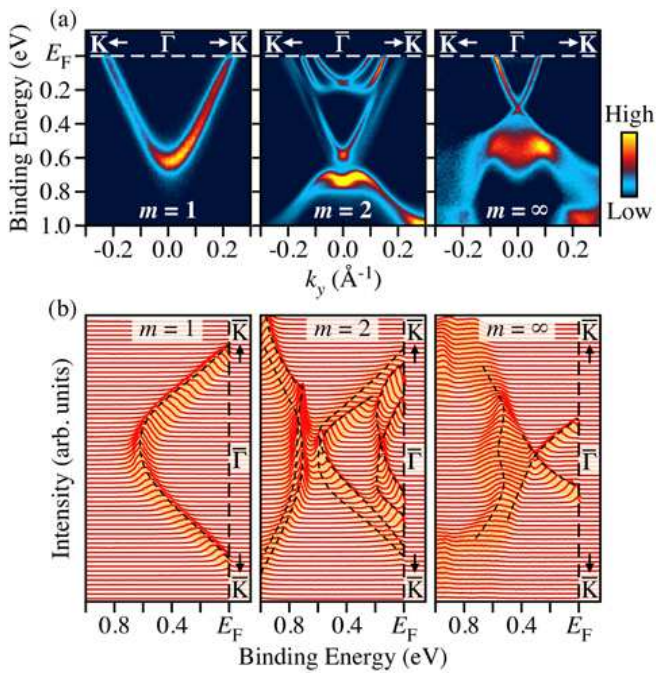}
\end{center}
\caption{
(color online). (a),(b) Comparison of near-$E_{\rm F}$ ARPES intensity (a) and corresponding energy dispersion curves (b). Dashed curves in (b) are a guide to the eyes.
}
\end{figure}

A side-by-side comparison of the band dispersions near $E_{\rm F}$ in Figs. 2(a) and 2(b) clearly shows a marked difference in $m$ = 1, 2, and $\infty$. Specifically, the band structure for $m$ = 1 contains a single, parabolic electron-like band, while that for $m$ = 2 is composed of multiple bands. The band structure at $m$ = 2 is also distinct from that at $m$ = $\infty$ where a simple ``X"-shaped Dirac-cone surface state shows up. We argue that such a striking difference in the near-$E_{\rm F}$ electronic states originates from the difference in the number of Bi$_2$Se$_3$ QLs in a unit cell, which is inferred from the fact that the ARPES data for $m$ = 1 and 2 bear spectral signatures resembling those of Bi$_2$Se$_3$ ultrathin films \cite{XueNP}; namely, the $m$ = 1 data show only the upper parabola, as was the case for 1-QL film, and the $m$ = 2 data present a gap with the lower band having a shallow dip at the top, similarly to the case for 2-QL film. This strongly suggests that the electronic states for $m$ = 1 and 2 are quantized due to electron confinement within the Bi$_2$Se$_3$ layer, which would be expected in view of the insulating nature of the PbSe block layer indicated by the absence of any additional $E_{\rm F}$-crossing bands.

\begin{figure}[!t]
\begin{center}
\includegraphics[width=3.4in]{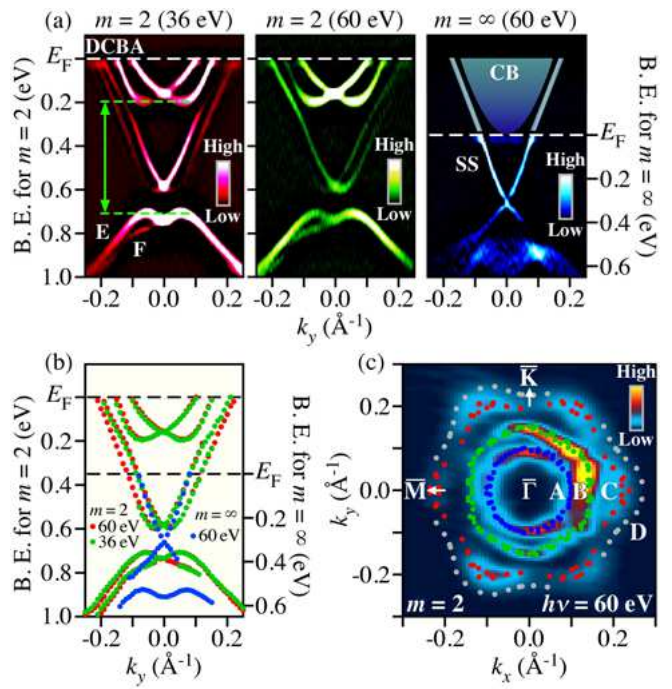}
\end{center}
\caption{
(color online). (a) Second-derivative ARPES intensities near the $\bar{\Gamma}$ point for $m$ = 2 plotted as a function of $k_y$ and $E_{\rm B}$ measured at $h$$\nu$ = 36 (left) and 60 eV (center), compared to that for $m$ = $\infty$ (right, measured with $h$$\nu$ = 60 eV) which is shifted downward by 0.35 eV to take into account the doping difference. Green arrows and dashed lines depict a conservative estimate of the bulk-band gap of 0.5 eV, judged from the energy difference between the edges of the bands B and E; notice that the band E is likely of the surface origin and hence the true bulk-band gap may well be larger than 0.5 eV. The band dispersions expected for $m$ = $\infty$ above $E_{\rm F}$ are also illustrated. (b) Direct comparison of the band dispersions obtained from the peak positions in the energy distribution curves for $m$ = 2 and $\infty$; the dispersions for $m$ = $\infty$ are shifted down to account for the doping difference. (c) ARPES intensity mapping at $E_{\rm F}$ for $m$ = 2 plotted as a function of 2D wave vector; this intensity is obtained by integrating the spectra within $\pm$10 meV of $E_{\rm F}$. The Fermi vector of each pocket is plotted by dots.
}
\end{figure}

To gain further insights into the origin of the difference between $m$ = 1 and 2, it is useful to examine the $m$ = 2 data in detail. As shown in Figs. 3(a) and 3(b), the near-$E_{\rm F}$ dispersions for $m$ = 2 are composed of six dispersive features labeled A-F, all of which show no photon-energy dependence signifying negligible dispersion along the momentum perpendicular to the surface ($k_z$) and hence their two-dimensional (2D) nature. The electron-like bands A and B cross at the $\bar{\Gamma}$ point at $E_{\rm B}$ = 0.16 eV and form two concentric Fermi surfaces [A and B in Fig. 3(c)], which indicates that their origin is the quantized bulk conduction bands with Rashba splitting \cite{KingRashba, BeniaRashba}. The dispersions of bands C and D in Fig. 3(a) are similar to those of A and B, but they should be attributed to the topological Dirac-cone states because (i) their overall dispersions overlap with the Dirac-cone surface states for $m$ = $\infty$ [see Fig. 3(b)] and (ii) the corresponding Fermi surfaces [C and D in Fig. 3(c)] exhibit a hexagonal deformation similarly to the case of $m$ = $\infty$ \cite{ChenBi2Te3, KurodaWarping}. It is therefore most natural to interpret that the bands C and D originally arise as topological interface states of the 2-QL Bi$_2$Se$_3$ unit at its top and bottom, but they are hybridized to open a gap \cite{QPTTheory1, QPTTheory2} and further are Rashba-split. In fact, they smoothly connect to bands E and F at the higher $E_{\rm B}$ side (see the schematic in Fig. 5). Note that the Rashba splitting in the upper branch of the hybridized topological states has already been reported in ultrathin films \cite{XueNP}. The present result is probably the first observation of the existence of topological states at the heterostructure ${\it interface}$ of a bulk crystal.

\begin{figure}[!t]
\begin{center}
\includegraphics[width=3.4in]{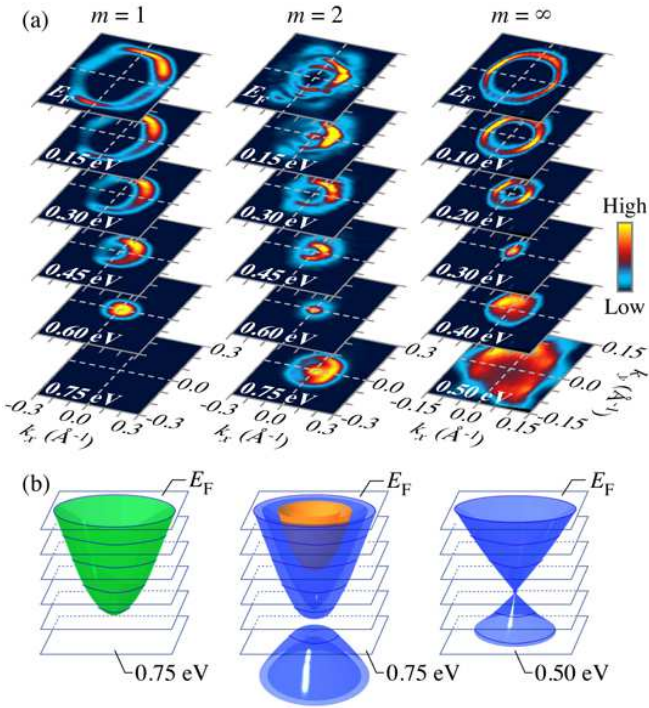}
\end{center}
\caption{
(color online). (a) ARPES intensity mappings for $m$ = 1, 2, and $\infty$ as a function of 2D wave vector at various $E_{\rm B}$'s. (b) Schematic 3D plots of the near-$E_{\rm F}$ band dispersions showing the energy positions for which the 2D intensity mappings in (a) are displayed.
}
\end{figure}

The next important issue is the origin of the marked difference between $m$ = 1 and 2. As discussed above, the 2D states C, D, E, and F observed for $m$ = 2 are most likely of topological origin. In contrast, the parabola observed near $E_{\rm F}$ for $m$ = 1 does not have a corresponding lower branch, which can be also seen in the iso-energy cuts of the spectra shown in Fig. 4, suggesting that its origin is distinct from a gapped Dirac cone. In ref. 21, the absence of the lower branch in the 1-QL film was proposed to be due to bonding with the substrate; similarly, we speculate that the standard model Hamiltonian \cite{ModelHamiltonian} is not applicable to 1 QL of Bi$_2$Se$_3$ sandwiched by PbSe layers and the system is in a topologically trivial state (similar conclusion was also proposed in ref. 28). If this is actually the case, the difference between $m$ = 1 and 2 signifies a topological quantum phase transition (QPT) and the observed parabola for $m$ = 1 is the degenerate lowest subband of the quantum confined conduction band.

\begin{figure}[!t]
\begin{center}
\includegraphics[width=3.4in]{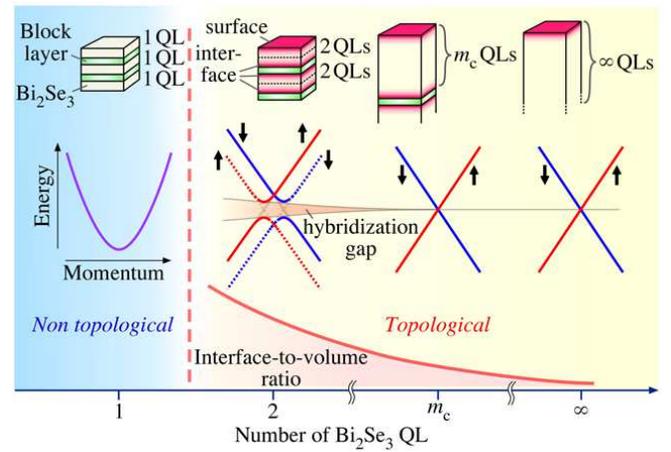}
\end{center}
\caption{
(color online). Schematic band diagrams and illustrations of the heterostructures for varying $m$. The red area in heterostructures depicts the topological interface (surface) states. $m_c$ is the characteristic thickness where the hybridization gap between two topological interface states starts to open. The interface-to-volume ratio is also shown schematically at the bottom.
}
\end{figure}

Figure 5 summarizes the evolution of the near-$E_{\rm F}$ electronic structure upon varying $m$, highlighting the putative topological QPT. The hybridization gap in the Dirac-cone interface states observed on the topological side is gradually reduced with increasing $m$, and it disappears at a characteristic $m$ value, $m_c$, above which the coherent coupling between two adjacent interface states becomes negligibly weak. One can view this evolution as a consequence of the finite size effect in TI slabs embedded in naturally-occurring heterostructures.

Perhaps more importantly, our results bear important implications on the band-gap engineering of TIs. Namely, as shown in Fig. 3, the magnitude of the bulk-band gap for $m$ = 2 is at least 0.5 eV, which is much larger than that for $m$ = $\infty$ ($\sim$0.3 eV) \cite{HasanBi2Se3} and is the largest ever observed in TIs. This demonstrates that the bulk-band gap can be enhanced by taking advantage of the quantization of the bulk bands in heterostructures, which is very useful for realizing TI devices working at room temperature.

It is useful to note that the existence of topological 2D states at the interfaces contained in naturally-occurring heterostructures has a practical importance. It is well known that the surface of 3D TIs is prone to chemical reactions in ambient atmosphere \cite{TaskinPRL, KongNChem, BeniaRashba}, which has hampered transport studies of topological surface states that are prerequisite to device applications, calling for an efficient means to protect the surface from degradation. Obviously, the topological interfaces found here reside in bulk crystals and hence are efficiently protected from ambient atmosphere. Moreover, those interface states naturally lead to sizable interface-to-volume ratio (see Fig. 5), which would boost the possibility of utilizing the topological 2D states in bulk crystals.

In conclusion, our ARPES measurements of (PbSe)$_5$(Bi$_2$Se$_3$)$_{3m}$ revealed a Rashba splitting of topological 2D states for $m$ = 2, providing the first experimental evidence for the existence of topological states at the heterostructure interface in the bulk. The natural multilayer structure leads to the protection of topological interface states and sizable interface-to-volume ratio. We also found that the band dispersions of Bi$_2$Se$_3$ layers are quantized due to confinement effects and exhibit a drastic change upon varying $m$. In particular, the largest bulk-band gap ever observed in a TI is achieved at $m$ = 2. Furthermore, we observed an intriguing topological phase transition between $m$ = 1 and 2. Those results demonstrate that naturally-occurring heterostructures are a promising playground for realizing novel topological phenomena and device applications of TIs.

We thank T. Arakane, M. Komatsu, K. Yoshimatsu, H. Kumigashira, and K. Ono for their assistance in ARPES measurements. This work was supported by JSPS (NEXT Program, KAKENHI 23224010, and Grant-in-Aid for JSPS Fellows 23.4376), JST-CREST, MEXT of Japan (Innovative Area ``Topological Quantum Phenomena"), AFOSR (AOARD 124038), and KEK-PF (Proposal number: 2012S2-001).

\end{document}